\documentclass[aps, prd]{revtex4-2}%
\usepackage{eurosym}
\usepackage{amsmath,amssymb}
\usepackage{graphicx}
\usepackage{dcolumn}
\usepackage{bm}
\usepackage{color}
\usepackage{multirow}
\usepackage{float}
\usepackage{array}
\usepackage{dcolumn}
\usepackage{booktabs}
\usepackage[detect-all]{siunitx}
\usepackage{makecell}
\usepackage{xcolor}
\usepackage{yfonts}

\newcolumntype{P}[1]{>{\centering\arraybackslash}p{#1}}
\graphicspath{{./}}

\begin{document}

\title{On $\Omega_{3c}NN$ and $\Omega_{3c} \Omega_{3c} N$ systems with HAL QCD potentials}
\author{Igor Filikhin$^1$, Roman Ya. Kezerashvili$^{2,3,4}$, and Branislav Vlahovic$^1$}
\affiliation{\mbox{$^{1}$North Carolina Central University, Durham, NC, USA} \\
$^{2}$New York City College of Technology, The City University of New York,
Brooklyn, NY, USA\\
$^{3}$The Graduate School and University Center, The City University of New
York, New York, NY, USA\\
$^{4}$Long Island University, Brooklyn, NY, USA}

\begin{abstract}
\noindent
This study employs the Faddeev formalism in configuration space to investigate the
$\Omega_{3c}NN$ cluster containing a triply charmed Omega baryon ($\Omega_{3c}$).
Using the recently reported HAL QCD $S$-wave $\Omega_{3c}N$ potentials in the
$^3S_1$ and $^5S_2$ channels, together with the MT-I--III nucleon--nucleon potential
and neglecting the Coulomb force, we find no bound state for the $\Omega_{3c}np$ system.
We predict near-threshold resonances in the $J^{\pi}=5/2^{+}$ (maximal total spin) and $J^{\pi}=1/2^{+}$ (minimal total spin)
states, with resonance energies of $1.1~\mathrm{MeV}$ below and $0.0~\mathrm{MeV}$ at the three-body breakup threshold, respectively, at Euclidean time $t/a = 16$.
A similar analysis of the $\Omega_{3c}\Omega_{3c}N$ system likewise reveals no bound states, though a possible resonance is indicated. The short-distance behavior of the HAL QCD
$\Omega_{3c}N$ potential is also discussed.

\end{abstract}

\maketitle


\noindent \textbf{Introduction.} In their seminal 1977 paper~\cite{PhysRevLett.39.1506}, Dover and Kahana first proposed the existence of charmed hypernuclei, predicting that $\Lambda_c$ baryons could be bound in nuclei through attractive nuclear interactions. In a follow-up study with Trueman~\cite{PhysRevD.16.799}, they extended this idea to include both charmed and anticharmed baryons, showing that while $\Lambda_c$ states could be stable, anticharmed systems would be limited by annihilation effects. Together, these pioneering works provided the theoretical foundation for later investigations of exotic nuclear systems, influencing both hadronic structure studies and future experimental programs in high-energy nuclear and particle physics.

The study of dibaryons is one of the long-standing problems in hadron physics. The long history of dibaryon searches, with attempts dating back to the 1950s, has been marked by many ups and downs and is reviewed in Ref.~\cite{Clement2017}. On the theoretical side, the $N\Omega$ state with strangeness $-3$ has been investigated by several groups~\cite{Goldman1987,Oka1988,Wang1995,Li2000,Pang2004,Chen2011,Huang2015,Morita2016,Sekihara2018,Iritani2019}. Tribaryon clusters such as $\Omega NN$ with strangeness $-3$ have also been studied within few-body approaches employing either the HAL QCD interaction~\cite{Iritani2019} or local meson-exchange potentials~\cite{Sekihara2018}, as reported in Refs.~\cite{Garcilazo2018,Garcilazo2019,Zhang2022,Etminan2023,Filikhin2025OmegaSSS}. Both types of interactions indicate the existence of the $\Omega N$ bound state and lead to a bound $\Omega NN$ tribaryon with strangeness~$-3$.

The most strange dibaryon, $\Omega_{3s}\Omega_{3s}$, has been studied on the basis of $(2+1)$-flavor lattice QCD simulations with a large volume and nearly physical pion mass, $m_\pi \simeq 146~\text{MeV}$~\cite{Gongyo2018}. The results indicate that the $\Omega_{3s}\Omega_{3s}$ system exhibits an overall attraction that leads to the $^1S_0$ bound state. Studies of the $\Omega_{3s}\Omega_{3s}N$ system using HAL QCD interactions~\cite{Iritani2019,Gongyo2018}, as well as investigations of exotic hypernuclei such as $^5_{\Omega}\text{He}$, $^6_{\Omega\Omega}\text{He}$, and $^9_{\Omega}\text{Be}$—treated as $\Omega\alpha$, $\Omega\Omega\alpha$, and $\Omega\alpha\alpha$ systems, respectively, within two- and three-particle cluster models—have also been performed \cite{Kezerashvili2024APS,Kezerashvili2024Fold,Etminan2025Omega}. These studies, employing an $\Omega\alpha$ interaction constructed from the HAL QCD $\Omega N$ potential in the $5/2$ spin channel together with the recently suggested $\Omega_{3s}\Omega_{3s}$ HAL QCD interaction~\cite{Gongyo2018}, demonstrate the possible formation of such hypernuclei.

In Ref.~\cite{Bjorken1985}, it was pointed out that baryons composed of three charm quarks could provide a novel perspective on baryon structure and spectroscopy. More recently, the authors of Ref.~\cite{Huang2020} investigated the possibility of heavy-quark dibaryons, focusing on the $\Omega_{3c}N$ system, by employing the Chiral Quark Model 
and the Quark Delocalization Color Screening Model. 
Using the resonating group method with both color-singlet and hidden-color channel couplings, they systematically analyzed the interaction between a nucleon and a triply charmed baryon. A first lattice QCD simulation of the $\Omega_{3c}N$ potentials in the $^3S_1$ and $^5S_2$ channels at the physical pion mass ($m_\pi= 137.1$ MeV) has recently been published by the HAL QCD Collaboration~\cite{Zhang-NOmegaccc}, where overall attraction was found in both channels. Furthermore, a pair of triply charmed baryons, $\Omega_{3c}\Omega_{3c}$, has been studied as an ideal dibaryon system in $(2+1)$-flavor lattice QCD with nearly physical light-quark masses and the relativistic heavy-quark action for the physical charm quark mass~\cite{Lyu2021_O_O}. In addition, Ref.~\cite{Huang2020ccc} suggests the possible existence of a dibaryon $\Omega_{3c}\Omega_{3c}$ with charm $C=6$. Thus, both $\Omega_{3c}N$ and $\Omega_{3c}\Omega_{3c}$ interactions appear to be attractive. As a natural extension, the investigation of charmed tribaryon systems such as $\Omega_{3c}NN$ and $\Omega_{3c}\Omega_{3c}N$ becomes of great interest.

In this letter, we report the study of $\Omega_{3c}NN$ and $\Omega_{3c}\Omega_{3c}N$ tribaryons with charm $3$ and $6$, respectively, within the framework of the Faddeev equations in configuration space~\cite{Faddeev1961,Faddeev1963,Noyes1968,Gignoux1974,Faddeev1993}, employing the HAL QCD $\Omega_{3c}N$~\cite{Zhang-NOmegaccc} and $\Omega_{3c}\Omega_{3c}$~\cite{Gongyo2018} interactions. We predict near-threshold resonances for the $J^{\pi}=5/2^{+}$ (maximal total spin) and $J^{\pi}=1/2^{+}$ (minimal total spin) states, with resonance energies of $-1.1~\mathrm{MeV}$ below and $0.0~\mathrm{MeV}$ at the three-body breakup threshold, respectively. 
In the $\Omega_{3c}\Omega_{3c}N$ system, a resonance is likely to exist; however,
it appears to be located well above the three-body threshold.


\noindent\textbf{Interactions.} \label{sec:Two-body-potentials} A study of $\Omega_{3c} NN$ and $\Omega_{3c} \Omega_{3c} N$ system with charm 3 and 6, respectively, within a nonrelativistic potential model using the Faddeev formalism requires $NN$, $\Omega_{3c} N$, and $\Omega_{3c}\Omega_{3c}$ interaction potentials.

\noindent\textbf{\textit{$NN$ interaction.}}
We employ the same $NN$ MT-I-III potential \cite{Malfliet1969,MTcorr} that has been used in studies of tribaryons with strangeness $0$ and $-3$, such as $\Omega_{3s} NN$, in Refs.~\cite{Garcilazo2018,Garcilazo2019,Zhang2022,Etminan2023,Filikhin2025OmegaSSS}.
It should be noted that the MT-I-III potential provides a good description of the low-energy $NN$ parameters and successfully reproduces the properties of three- and four-nucleon systems within an $S$-wave model.

\noindent\textbf{\textit{$\Omega N$ interaction.}}
Most recently, the authors of Ref.~\cite{Zhang-NOmegaccc} investigated the
$\Omega_{3c} N$ system to explore the interaction between a nucleon and a heavy charm baryon using
the HAL QCD method~\cite{Ishii2007,Ishii2013,Aoki2020},
a non-perturbative framework that determines interactions directly from
spacetime correlation functions of multiple hadrons.
Note that among the various dihadron systems studied with the time-dependent HAL QCD method are $\Omega_{3s}N$ in the $^5S_{2}$ channel~\cite{Iritani2019},
$\phi N$ in both the $^4S_{3/2}$~\cite{Lyu2022} and $^2S_{1/2}$~\cite{Chizzali2024} channels,
and $N$--$J/\psi$ and $N$--$\eta_{c}$ in both the $^4S_{3/2}$ and $^2S_{1/2}$ channels~\cite{Lyu2025}, each exhibiting distinct quark-flavor content.

In Ref.~\cite{Zhang-NOmegaccc} the $\Omega_{3c}N$ system is investigated in both $^3{\rm S}_1$ and $^5{\rm S}_2$ channels to explore the interaction between a nucleon and a heavy baryon using $(2+1)$-flavor lattice QCD with the lattice spacing is $a \simeq  0.084372$ fm, yielding a sufficiently large spatial extent
of $L \simeq  8.1$ fm to accommodate two-baryon systems. Simulations were performed with
nearly physical quark masses, corresponding to a pion mass of $m_\pi =137.1$ MeV  and a physical charm quark mass defined by
$\left(m_{\eta_c} + 3m_{J/\psi}\right)/4 \simeq 3068.5\,\text{MeV}$. The masses of the spin-averaged charmonium
state $m_{\Omega_{3c}}$ for baryon calculated with single exponential fits over the interval $t/a = 30 - 40$ are 4846.4 MeV and 4846.4 MeV. The interpolated mass is 4796.8 MeV.

Because the masses of the $\Lambda_{c}\Xi_{3c}$ and $\Sigma_{c}\Xi_{2c}$ baryons lie above the $\Omega_{3c}N$ threshold, a reliable extraction of the low-energy $N\Omega_{3c}$ interaction in both the $^3S_{1}$ and $^5S_{2}$ channels is possible via the time-dependent HAL QCD calculations. The extracted potential was fit using a phenomenological two-range Gaussian form \cite{Zhang-NOmegaccc}
	\begin{equation}
		V_{\Omega_{3c} N}\left(r\right)=\alpha_{1}e^{-\left(r/\beta_{1}\right)^{2}}+\alpha_{2}e^{-\left(r/\beta_{2}\right)^{2}}.  \label{eq:pot_nO}
	\end{equation}
The parameters for the uncorrelated fits in the range $ 0 < r < 3 $ fm for both the $^3S_{1}$ and $^5S_{2}$ channels are given in Table~\ref{tab:par}.
The interaction (\ref{eq:pot_nO}) is attractive in both channels, with scattering lengths
$a_0 \simeq 0.56~\text{fm}$ and $a_0 \simeq 0.38~\text{fm}$, and effective ranges
$r_{\rm eff} \simeq 1.60~\text{fm}$ and $r_{\rm eff} \simeq 2.04~\text{fm}$, respectively.
For the HAL QCD interaction (\ref{eq:pot_nO}), no bound state is observed in the $\Omega_{3c}N$ system in lattice QCD results~\cite{Zhang-NOmegaccc}, in contrast to some predictions from the quark model~\cite{Huang2020}.

\begin{figure}[ht]
\includegraphics[width=20pc]{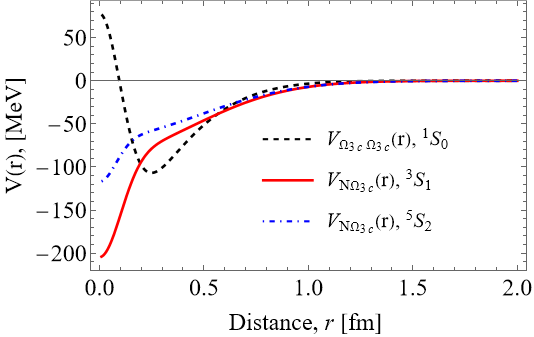}
\caption{ HAL QCD $\Omega_{3c}\Omega_{3c}$ potential (dashed line) in channel $^1S_0$ and $N\Omega_{3c}$ interactions in channels $^3S_1$ (solid line) and $^5S_2$, respectively (dot-dashed line). }
\label{fig:1}
\end{figure}		
		
   \begin{table}[htp]

\caption{\label{tab:merged} Fitting parameters for $N\Omega_{3c}$ and $\Omega_{3c}\Omega_{3c}$ potentials. The numbers in parentheses indicate statistical errors. Parameters are taken from Refs.~\cite{Zhang-NOmegaccc,Lyu2021_O_O}.}

\centering

\begin{tabular}{lccccccc}

\noalign{\smallskip}\noalign{\smallskip}
\hline
\noalign{\smallskip}
\multicolumn{8}{c}

{$\Omega_{3c}N$ potential in the $^3S_1$ and $^5S_2$ channels at various lattice Euclidean time $t/a$} \\

\noalign{\smallskip}\hline  \noalign{\smallskip}

Channel & $t/a$ & $\alpha_1$ (MeV) & $\beta_1$ (fm) & $\alpha_2$ (MeV) & $\beta_2$ (fm) & & \\

\noalign{\smallskip}\hline
\noalign{\smallskip}
 & 16 & -118.9(1.6) & 0.142(7) & -85.7(2.6) & 0.633(33) & & \\

$^{3}S_{1}$ & 17 & -118.0(3.0) & 0.135(8) & -80.0(3.7) & 0.601(37) & & \\

& 18 & -119.7(4.6) & 0.141(12) & -75.0(8.5) & 0.608(56) & & \\

\noalign{\smallskip}\hline
\noalign{\smallskip}
&  16 & -50.5(3.6) & 0.110(15) & -66.5(3.0) & 0.665(46) & & \\

$^{5}S_{2}$ & 17 & -52.6(2.5) & 0.110(12) & -60.4(2.5) & 0.612(50) & & \\

& 18 & -53.0(5.2) & 0.113(21) & -57.8(5.2) & 0.636(67) & & \\

\noalign{\smallskip}\hline
\noalign{\smallskip}
\multicolumn{8}{c}{$\Omega_{3c}\Omega_{3c}$ potential in the $^1S_0$ channel at tlattice Euclidean time $t/a=26$} \\

\noalign{\smallskip}\hline
\noalign{\smallskip}
Channel & $\alpha_1$ (MeV) & $\beta_1$ (fm) & $\alpha_2$ (MeV) & $\beta_2$ (fm) & $\alpha_3$ (MeV) & $\beta_3$ (fm) \\

\noalign{\smallskip}\hline
\noalign{\smallskip}
$^{1}S_{0}$ & 239.(3.0) & 48.5(1.4) & $-62.7(50.8)$ & 7.8(2.6) & $-98.8(50.3)$ & 3.4(0.8) \\

\noalign{\smallskip}\hline
\label{tab:par}
\end{tabular}

\end{table}

\noindent\textbf{\textit{$\Omega_{3c}\Omega_{3c}$ interaction.}}
In Ref.~\cite{Lyu2021_O_O}, the first investigation of the scattering properties of the
$\Omega_{3c}\Omega_{3c}$ system was presented on the basis of
$(2+1)$-flavor lattice QCD simulations with the physical charm quark mass and nearly physical light quark masses.
The potential for $\Omega_{3c}\Omega_{3c}$ in the $^{1}S_{0}$ channel was obtained by the time-dependent HAL QCD method without including the Coulomb interaction. 
In the range $r \leq 2.5~\mathrm{fm}$ this potential was fitted by a sum of three-range Gaussians
\begin{equation}
		V_{\Omega_{3c}\Omega_{3c}}\left(r\right)=\sum_{i=1}^{3}\alpha_{i}e^{-\beta_{i}r^2},
        \label{eq:pot_OmegacccOmegaccc}
	\end{equation}
with the fitting parameters listed in Table.~\ref{tab:par} at Euclidean time $t/a = 26$. The potentials are found at Euclidean times $t/a = 25, 26,$ and $27$ to be consistent with each other within the statistical errors~\cite{Lyu2021_O_O}.
The potential (\ref{eq:pot_OmegacccOmegaccc}) exhibits short-range repulsion and
intermediate-range attraction, consistent with the qualitative behavior of
the $NN$ interaction and the $\Omega_{3s}^{-}\,\Omega_{3s}^{-}$ potential~\cite{Gongyo2018}.
The comparison of the $\Omega_{3s}\Omega_{3s}$ and $\Omega_{3c}\Omega_{3c}$ potentials shows that $V_{\Omega_{3c}\Omega_{3c}}$ is much deeper at short distances and more strongly repulsive than $V_{\Omega_{3s}\Omega_{3s}}$, while the latter is much wider.
Notice that the masses obtained, $m_{\Omega_{3c}} = 4796.8(0.1)$~\cite{Zhang-NOmegaccc} and $m_{\Omega_{3c}} = 4795.6(0.7)$~\cite{Lyu2021_O_O}, are consistent with $4789(6)(21)$~MeV obtained from $(2+1)$-flavor configurations generated by the PACS-CS Collaboration~\cite{Namekawa2013}.

The HAL QCD $\Omega_{3c}\Omega_{3c}$ potential in the $^1S_0$ channel, together with the $N\Omega_{3c}$ potential in the $^3S_1$ and $^5S_2$ channels, are shown in Fig.~\ref{fig:1}. A comparison of these potentials indicates that the $\Omega_{3c}\Omega_{3c}$ potential exhibits a repulsive core at short distances, in contrast to the $\Omega_{3c}N$ potential in the $^3S_1$ and $^5S_2$ channels.  At the same short-range separation between the $\Omega_{3c}$ baryon and the nucleon, the interaction in the $^3S_1$ channel is nearly twice as attractive as that in the $^5S_2$ channel.

\noindent\textbf{Theoretical approach.} $\Omega_{3c} NN$ and $\Omega_{3c} \Omega_{3c} N$ system with charm 3 and 6 presnts a three-body system with two identical particles and can be investigated in the framework of the Faddeev formalism. The Faddeev equations in configuration space
for three-body systems with two identical bosons or fermions are given in our previous
studies \cite{Kez2017,Kez2018PL,KezerasPRD2020}. In the case of two identical fermions,  one must account for the antisymmetrization of the total wave function, and the total wave function of the system is decomposed into the sum
of the Faddeev components $\Phi_1$ and $\Phi_2$ corresponding to the $\Omega_{3c} (NN)$ [$ (\Omega_{3}\Omega_{3c})N$] and $(\Omega_{3c} N)N$ [$\Omega_{3c}(\Omega_{3c}N)$] types of rearrangements for $\Omega_{3c} NN$ and $\Omega_{3c} \Omega_{3c} N$, respectively: $\Psi =\Phi_1+\Phi_2-P\Phi_2$,
where $P$ is the permutation operator for two identical fermions.
Finally, the set of the Faddeev equations for a system with two identical particles reads
\cite{Kvitsinsky1986,Kvitsinsky1986Sov}:
\begin{equation}
\begin{array}{l}
{(H_{0}+V_{AA}-E)\Phi_1=-V_{AA}(\Phi_2-P\Phi_2)}, \\
{(H_{0}+V_{AB}-E)\Phi_2=-V_{AB}(\Phi_1-P\Phi_2)}.%
\end{array}
\label{GrindEQ__1_}
\end{equation}
In Eqs.~(\ref{GrindEQ__1_}), $V_{AA}$ denotes the interaction potential between identical particles,
either nucleons ($V_{NN}$) or baryons ($V_{\Omega_{3c}\Omega_{3c}}$),
while $V_{AB}$ describes the interaction between nonidentical particles,
namely the $\Omega$ baryon and the nucleon ($V_{\Omega N}$).

The spin--isospin variables of the system can be represented by the corresponding bases
$\xi_{\text{isospin}}$ and $\eta_{\text{isospin}}$.
One can then define the coordinate part of the total wave function, $\Psi^{R}$, as follows:
$\Psi = \xi_{\text{isospin}} \otimes \eta_{\text{isospin}} \otimes \Psi^{R}$.
We consider an $S$-wave model in which  the total orbital angular momentum
and the pair orbital angular momenta are equal to zero.
\noindent\textbf{Numerical Results and Discussion.}
In the calculations, we use the potentials $\Omega_{3c}N$ \cite{Zhang-NOmegaccc} and $\Omega_{3c}\Omega_{3c}$ \cite{Lyu2021_O_O} with the parameters listed in Table~\ref{tab:par}, corresponding to Euclidean times $t/a = 16-18$ and $t/a = 26$, respectively, together with the $\Omega_{3c}$ mass of 4796~MeV/$c^2$.
Our calculations confirm the result of \cite{Zhang-NOmegaccc} that the $\Omega_{3c}N$ two-body system in the $ ^{3}S_{1}$ and $ ^{5}S_{2} $ channels is not bound. The strong attractive interaction in the channels produces too shallow and narrow quantum well.
\begin{table}[htp]
		\caption{
		The resonance energy in MeV of the $\Omega_{3c}NN$ system
in the state $(T)J^{\pi}$.
The $\Omega_{3c}N$ potential was taken at various lattice Euclidean times,
$t/a = 16, 17,$ and $18$.
The energy is measured relative to the three-body breakup threshold.
}
\begin{tabular}{ccc}
\noalign{\smallskip}\noalign{\smallskip}\hline
$ t/a$ & $(0)1/2^{+}$ &  $(0)5/2^{+}$\\ \noalign{\smallskip}\hline
16 & -1.1 & 0.0 \\
17 & -0.7 & -0.5 \\
18& 0.1 & 0.2 \\ \noalign{\smallskip}\hline
\end{tabular}
\label{tab:1}
		\end{table}
Our calculations within the framework of the Faddeev equations for the
$\Omega_{3c}np$ system, using the HAL QCD $\Omega_{3c}N$ potential~\cite{Zhang-NOmegaccc},
indicate that no bound state is formed in the $J^{\pi} = 1/2^{+}$ and $5/2^{+}$ states
for Euclidean times $t/a = 16, 17,$ and $18$ of the $\Omega_{3c}N$ potential.

For comparison, in the $\phi N$ system, the HAL QCD potential in the
$^4S_{3/2}$ channel~\cite{Lyu2022} does not produce a bound $\phi N$ state~\cite{Lyu2022},
nor does it lead to a bound $\phi NN$ system~\cite{Filikhin2024phid,Filikhin2025Supatom}.
A bound state of the $\phi NN$ system, denoted as $^3_\phi$H,
is obtained only with the essentially stronger
$^2S_{1/2}$ potential~\cite{Chizzali2024}, as described in
Refs.~\cite{Filikhin2024phid,Filikhin2025Supatom,Wen2025}.

\begin{figure}[ht]
\includegraphics[width=20.4pc]{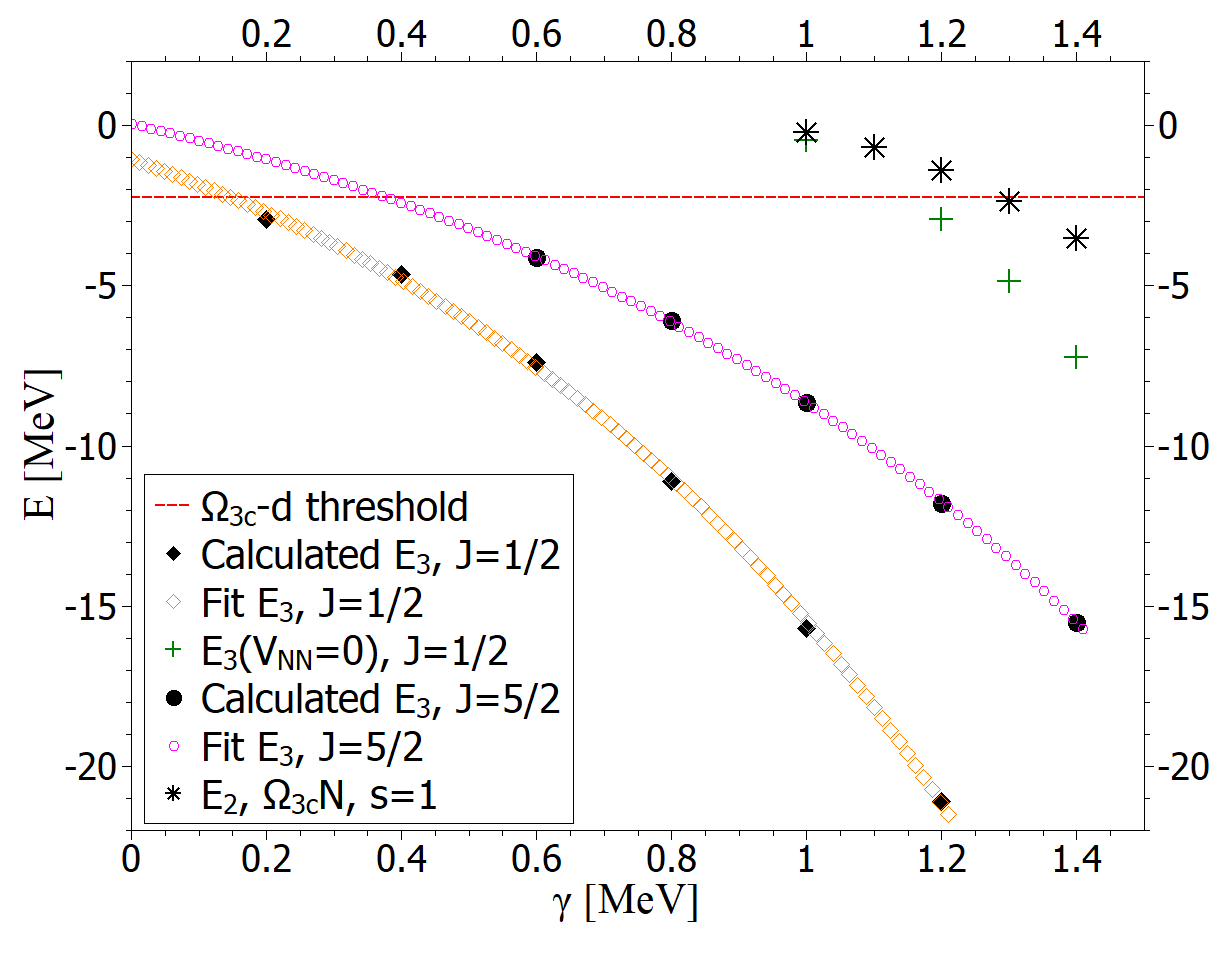}
\includegraphics[width=20pc]{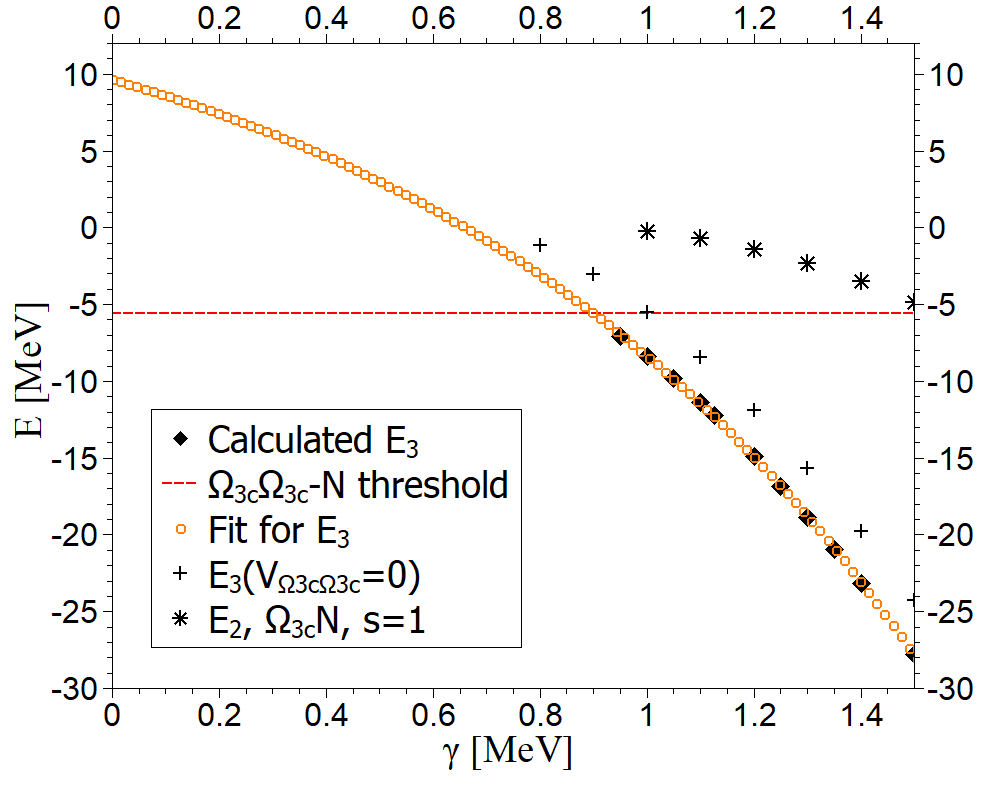}
\caption{\label{fig:1}
\textbf{\textit{Left panel:}} Evaluation of resonance energies for the $J^\pi=1/2^{+}$ and $5/2^{+}$ states of the $\Omega_{3c}np$ system.
Three-body energy $E_3$ is presented as a function of the scaling parameter $\gamma$.
The exponential interpolation (\ref{3}) 
is used to extrapolate the energies corresponding to the resonance positions
relative to the $\Omega_{3c}$-$d$ two-body breakup threshold (dashed line).
Solid diamonds and circles show results for the $\Omega N$ potential in
the $^3S_1$ and $^5S_2$ channels, respectively.
The resonance energy is defined at $\gamma = 0$.
Cross marks show energies $E_3$ calculated when interaction between two nucleons is neglected ($V_{NN}=0$),
and stars indicate the two-body binding energy of the $\Omega_{3c}N$ pair.
\textbf{\textit{Right panel:}}
Evaluation of the resonance energy in the $\Omega_{3c}\Omega_{3c}N$ system
with minimal quantum numbers $J^\pi = 1/2^+$.
The three-body energy $E_3$ is calculated for various scaling parameters $\gamma$
using the $\Omega N$ potential in the $^3S_1$ channel.
The $\Omega_{3c}\Omega_{3c}$-$N$ two-body breakup threshold is shown by the dashed line.
Calculated values of $E_3$ below the two-body breakup threshold are indicated by solid diamonds.
All other notations are the same as in the Left panel.
}
\end{figure}		
In the next step, we investigate the possible existence of resonances
in the $\Omega_{3c}NN$ system.
To estimate the resonance energies, we adopt an approach motivated by the method of
analytic continuation in the coupling constant proposed in Ref.~\cite{Kukulin}.
This technique has been successfully applied to few-body nuclear systems
\cite{Kurokawa,FKV}, where it has proven effective for identifying near-threshold
resonances and extracting the corresponding energy parameters.
By scaling the $\Omega_{3c}N$ potential as
\begin{equation}
\label{3}
V^{\gamma}_{\Omega_{3c}N} = V_{\Omega_{3c}N}(1+\gamma),
\end{equation}
where $\gamma$ is a fitting parameter that modifies the depth of the potential,
we obtain a bound state in the $\Omega_{3c}N$ system for fixed values of $\gamma$
in the $^{3}S_{1}$ and $^{5}S_{2}$ states. A series of calculations for different values of $\gamma$
yielded negative three-body energies $E_3$, which correspond to bound states of the $\Omega_{3c}np$ system.

To extend these results to positive energies, we employed an interpolation function of the form
$
f(\gamma) = A\, e^{\alpha \gamma} + B,
$
with constants $A$, $B$, and $\alpha$ determined from the calculated three-body energies $E_3$.
Uncertainties arise mainly from the choice of interpolation function and the sensitivity of the results
to variations in $\gamma$. A key factor in reducing interpolation errors is to bring the resonance closer
to the two-body $\Omega_{3c}d$ threshold. Figure~\ref{fig:1} presents the numerical results illustrating
the method used to evaluate the resonance energies for the $J^\pi=1/2^{+}$ and $5/2^{+}$ states.
The calculated resonance energies, summarized in Table~\ref{tab:1}, depend on the $t/a$ parameter of the potentials
and exhibit some degree of instability. This behavior arises from two main sources:
(i) uncertainties associated with the resonance-energy extraction, and
(ii) issues related to the stability of the potential calculations.

Motivated by the recently proposed HAL QCD potentials for the
$\Omega_{3c}\Omega_{3c}$ and $\Omega_{3c}N$ interactions with spins 0 and~1, respectively,
we also study the $\Omega_{3c}\Omega_{3c}N$ system.
The $\Omega_{3c}\Omega_{3c}$ interaction~\cite{Lyu2021_O_O} produces a bound dibaryon with charm $C=6$
and a binding energy of 5.54~MeV, consistent with the results of Ref.~\cite{Lyu2021_O_O}.
However, no bound state was found for the $\Omega_{3c}\Omega_{3c}N$ system.
Our results instead indicate the possible existence of a resonance located well above the three-body threshold.
An exponential interpolation, shown in Fig.~\ref{fig:1}, yields an approximate resonance energy of $\sim 9.5$~MeV,
although this value depends sensitively on the choice of interpolation function.
This sensitivity highlights the need for a more refined treatment of the interaction potentials
and more accurate methods for resonance-energy extraction.

In Fig.~\ref{fig:1}, we show the dynamics of the appearance of the bound state
in the spin-1 $\Omega_{3c}N$ system as a function of the scaling parameter $\gamma$.
This pair potential generates a bound state when the depth of the original potential
is increased by a factor of two. A similar behavior is observed for the three-body
$\Omega_{3c}\Omega_{3c}N$ system. The $\Omega_{3c}NN$ system is more sensitive to the
$\Omega_{3c}N$ potential, with a bound state  appearing for $\gamma \le 0.2$.
Comparison of the both systems indicates that the binding effect of the $NN$ potential
is stronger than that of the $\Omega_{3c}\Omega_{3c}$ potential.
This can be inferred from the differences between $E_3$ and $E_3(V_{NN}=0)$ for the $\Omega_{3c}NN$,
and between $E_3$ and $E_3(V_{\Omega_{3c}\Omega_{3c}}=0)$ for the $\Omega_{3c}\Omega_{3c}N$.
In the same Fig.~\ref{fig:1}, we show the dependence of the three-body energies
$E_3\big(V_{NN}=0\big)$ and $E_3\big(V_{\Omega_{3c}\Omega_{3c}}=0\big)$ on the parameter $\gamma$ when the interaction
between the identical baryons is ignored.
\begin{figure}[ht]
\includegraphics[width=15pc]{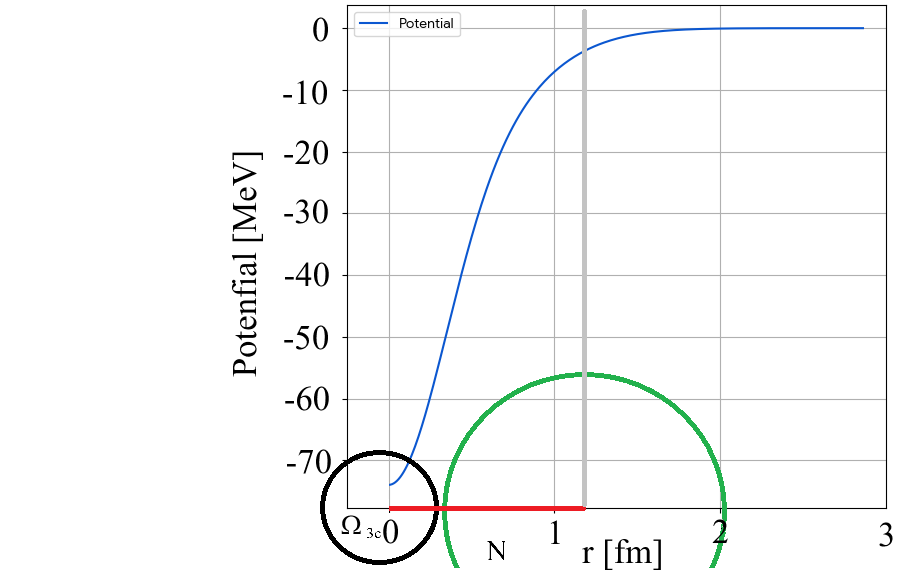}
\caption{\label{fig:01}
Modified $\Omega_{3c}N$ potential in the spin $^{3}S_{1}$ state that leads to the scattering length $a_{0} = 0.57~\text{fm}$ and effective range $r_{\text{eff}} = 1.77~\text{fm}$.  Circles indicate the approximate root-mean-square ($rms$) radii of the $\Omega_{3c}$ baryon and nucleon, positioned so the baryons just touch. The corresponding interparticle distance is shown by a red line.
}
\end{figure}		
	
The extracted $\Omega_{3}N$ interactions \cite{Zhang-NOmegaccc} are found to be attractive in both channels.
For the spin-1 channel, the HAL QCD potential $V_{\Omega N}$ yields a scattering length
$a_{0} = 0.56(13)$~fm and an effective range $r_{\text{eff}} = 1.60(5)$~fm~\cite{Zhang-NOmegaccc}.
For the purpose of a {\textit{discussion}, we considered a hypothetical two-range Gaussian potential
that reproduces similar low-energy parameters, namely
$a_{0} = 0.57~\mathrm{fm}$ and $r_{\text{eff}} = 1.77~\mathrm{fm}$.
Furthermore, the potential does not form a bound state
in the $\Omega_{3}N$ system.
This potential, shown in Fig.~\ref{fig:01}, is characterized by a substantially reduced short-range
attraction compared to the original HAL QCD interaction.
Obviously, the proposed potential primarily describes the $\Omega_{3}N$ interaction
at interparticle distances smaller than the nucleon $rms$  radius, $r_{\text{rms}}\approx 0.8$ fm,
which would correspond to the $\Omega_{3c}$ baryon penetrating into the nucleon.
This raises the question: can the $\Omega_{3c}$ reside inside a nucleon?

An $\Omega_{3c}$ baryon, composed of three charm quarks, cannot exist as a constituent inside a nucleon.
According to QCD, quarks are confined into color-singlet states, and embedding an $\Omega_{3c}$
within a nucleon would necessarily form a six-quark system, $uudccc$ or $uddccc$.
Such a configuration corresponds to an exotic multiquark state rather than a nucleon.
Furthermore, the mass of the $\Omega_{3c}$ baryon is expected to be much larger than that of a nucleon,
rendering such an embedded configuration energetically prohibited.

Instead, at this low energy the $\Omega_{3c}$ baryon can only interact with nucleons as a separate hadron,
with the possibility of forming weakly bound states or few-body systems such as
$\Omega_{3c}N$ or $\Omega_{3c}NN$.

The notion of placing one baryon ``inside'' another while preserving their individuality
is inconsistent with the theory of the strong interaction. Both the nucleon, $uud$ or $udd$, and the $\Omega_{3c}$ baryon ($ccc$) are color-singlet states,
meaning that their constituent quarks combine in such a way that the net color charge vanishes.
This singlet property is a fundamental requirement for any isolated hadron.
Because of color confinement, the strong interaction becomes increasingly strong as quarks are separated,
and forcing significant overlap between two color-singlet baryons does not yield a composite of two intact hadrons.
Instead, such overlap produces a six-quark configuration ($uudccc$), corresponding to an exotic multiquark state
with a complicated color structure. Such configurations are the subject of active research in hadron physics,
but they lie outside the scope of descriptions based on
baryon--baryon pair potentials. 

At short distances, the $\Omega_{3c}N$ interaction is expected to be characterized by a strong repulsive core.
This repulsion is not a consequence of the Pauli exclusion principle, since the quark flavors in $\Omega_{3c}$
and in the nucleon are distinct. Rather, it originates from the underlying QCD dynamics:
when the two baryons approach within their intrinsic sizes, the overlap of their color fields
leads to a rapid rise in potential energy. A color-singlet baryon cannot be smoothly deformed into
a highly overlapping configuration without excitation into higher-energy color states, which imposes
an energetic barrier. As a result, the formation of a compact six-quark configuration is strongly disfavored,
and an effective hard-core repulsion emerges, preventing the $\Omega_{3c}$ from penetrating the nucleon.


\noindent\textbf{Summary.}
Interactions in the charm sector based on lattice QCD provide a promising framework
for understanding heavy-baryon dynamics. Studies of tribaryon systems such as
$\Omega_{3c}NN$ and $\Omega_{3c}\Omega_{3c}N$, carried out within few-body approaches
like the Faddeev equations---the most rigorous method for describing three-particle systems---serve as
important tests for the $\Omega_{3c}N$ and $\Omega_{3c}\Omega_{3c}$ potentials.

We employed the Faddeev formalism in configuration space to study the $\Omega_{3c}np$
and $\Omega_{3c}\Omega_{3c}N$ tribaryons, which have charm numbers 3 and 6, respectively,
and include the deuteron and the bound dibaryon $\Omega_{3c}\Omega_{3c}$.
No bound state was found for the $\Omega_{3c}np$ system. However, we predict near-threshold resonances.
Specifically, using the HAL QCD potential with parameter $t/a = 1.6$, the $(T)J^{\pi} = (0)5/2^{+}$
and $(0)1/2^{+}$ states exhibit resonance energies of $1.1~\mathrm{MeV}$ below and $0.0~\mathrm{MeV}$
at the three-body breakup threshold, respectively.

A similar analysis was performed for the $\Omega_{3c}\Omega_{3c}N$ system, employing
the $\Omega_{3c}\Omega_{3c}$ and $\Omega_{3c}N$  potentials in singlet spin states.
No bound state was found in this case either, though the results suggest the existence
of a resonance located far above the three-body threshold.
The resonance energy was estimated to be approximately $9.5$~MeV, although this value
depends on the choice of approximation function used in the evaluation.

Finally, we analyzed the short-distance behavior of the HAL QCD potential for the
$\Omega_{3c}N$ interaction, where repulsive cores and internal quark-structure effects
are expected to play an important role.

\noindent{\bf Acknowledgement}			
This work is supported by the City University of New York, PSC CUNY Research Award Number 68541-00 56 and the Department of Energy/National Nuclear Security Administration Award Number NA0003979.

\end{document}